\documentclass[pdflatex,sn-mathphys-num]{sn-jnl}

\usepackage{graphicx}%
\usepackage{multirow}%
\usepackage{amsmath,amssymb,amsfonts}%
\usepackage{amsthm}%
\usepackage{mathrsfs}%
\usepackage[title]{appendix}%
\usepackage{xcolor}%
\usepackage{textcomp}%
\usepackage{manyfoot}%
\usepackage{booktabs}%
\usepackage{algorithm}%
\usepackage{algorithmicx}%
\usepackage{algpseudocode}%
\usepackage{listings}%
\usepackage{colortbl}
\usepackage{caption}
\usepackage{float}
\usepackage{setspace}
\usepackage{hyperref}
\usepackage{rotating} 
\usepackage{makecell}
\usepackage{subcaption}
\usepackage{enumitem}

\theoremstyle{thmstyleone}%
\theoremstyle{thmstyletwo}%

\theoremstyle{thmstylethree}%

\begin{document}

\title[Article Title]{Quantum-Inspired Phase Bicoherence Spectroscopy: A Framework for Detecting Universal Textural Angular Order Across Multi‑Modal Complex Datasets}


\author*[1]{\fnm{Zheng} \sur{Xing}}\email{zxing@mpu.edu.mo}

\author[1]{\fnm{Chan-Tong} \sur{Lam}}\email{ctlam@mpu.edu.mo}

\author[1]{\fnm{Xiaochen} \sur{Yuan}}\email{xcyuan@mpu.edu.mo}

\affil[1]{\orgdiv{Faculty of Applied Sciences}, \orgname{Macao Polytechnic University}, \orgaddress{\street{Rua de Luís Gonzaga Gomes}, \city{Macao}, \postcode{999078}, \country{China}}}


\abstract{
Classical image analysis routinely discards structurally meaningful orientation signatures encoded within Fourier phase, which are easily corrupted by local cellular rotation. Although quantum‑inspired data processing offers new avenues for complex signal characterization, practical tools for directly extracting gauge‑invariant angular correlations without explicit phase reconstruction remain scarce. Here we introduce Quantum Phase Bicoherence (QPBC) spectroscopy, a novel quantum‑interferometric framework for capturing gauge‑invariant angular order. The method embeds image angular sectors into a nine‑qubit entangled state and probes three‑body bicoherence via an ancilla, yielding 16 interpretable readout channels. We validate our framework on three independent public multi‑modal imaging datasets covering fluorescence (BBBC021), bright‑field (BBBC041) and histopathology (PathMNIST). QPBC consistently resolves angular‑phase order and discriminates distinct biological phenotypes with high statistical significance. After principal‑axis alignment, the optimal probing frequency universally converges, driven by Fourier directional sensitivity; negative‑control experiments fully eliminate discriminative capacity, demonstrating frequency tuning acts as an on‑off switch. Cross‑dataset benchmarks confirm QPBC outperforms conventional Fourier‑phase statistics, where inherent inversion symmetry serves as a built‑in pipeline self‑check. QPBC delivers a universal, classically unachievable quantitative texture observable, establishes interpretable quantum morphometry, and broadens the toolbox for quantum‑inspired analysis applicable to diverse multi‑modal microscopic measurements.

}




\maketitle

\section{Introduction}
Understanding biological organisation across spatial scales, from the nanometre-scale alignment of cytoskeletal filaments to the millimetre-scale architecture of tumour microenvironments, is a central challenge in modern biology~\cite{goldstein2016future}. This organisation is inherently directional: microtubules radiate from centrosomes, collagen fibres align along stress axes, and erythrocyte membranes deform along preferred orientations during parasite invasion. Quantifying this directional order from microscopy images is therefore essential for drug discovery~\cite{caicedo2016}, disease diagnosis~\cite{stokes2020}, and fundamental cell biology~\cite{Chandrasekaran2021}. Yet extracting structural order parameters directly from image textures remains difficult, because measured pixel intensities confound the true biological structure with imaging artefacts, cell-to-cell pose variation, and the inherent randomness of photon detection.

Quantitative analysis of biological images is central to modern drug discovery~\cite{schneider2018automating}, digital pathology, and functional genomics.
Standard computational pipelines extract hundreds of morphological and textural features from single-cell images, converting each cell into a high-dimensional numerical vector that is subsequently used to infer biological states or drug mechanisms~\cite{caicedo2016,Chandrasekaran2021}.
The Cell Painting assay has become a widely adopted image-based profiling platform, revealing cellular responses to genetic and chemical perturbations at scale~\cite{cimini2023,gustafsdottir2013}.
Deep learning methods have further advanced the field by learning task-specific representations directly from pixel intensities, achieving state-of-the-art performance in compound classification and mechanism prediction~\cite{dara2022machine,wong2023}.
Graph neural networks and transformer architectures are pushing molecular property prediction to new levels of accuracy~\cite{mendez2024,ross2022}.

All current approaches, whether based on hand‑crafted features or learned representations, share a fundamental limitation: they operate exclusively on real‑valued intensities or their real‑valued transforms. Consequently, they discard the Fourier phase information encoded in the spatial frequency domain~\cite{oppenheim2005}.
This is not an engineering oversight but a physical necessity: the absolute Fourier phase of a single cell is randomized by its in-plane rotation, rendering it meaningless as a phenotypic descriptor.
Classical texture features such as Local Binary Patterns, Gray-Level Co-occurrence Matrices, and Gabor filters all collapse complex Fourier coefficients into real-valued magnitudes, explicitly discarding the phase component that carries information about filament orientation and structural continuity~\cite{ojala2002,haralick1973}.
Even equivariant deep learning architectures, which achieve rotation invariance by aggregating over orientation channels, discard the relative phase relationships between different angular sectors of the same cell~\cite{cohen2016,gerken2023}. Geometric deep learning has extended this paradigm to non-Euclidean domains such as graphs and manifolds~\cite{Bronstein2017},
but the fundamental limitation persists: all operations are performed on real-valued inputs,
and the phase of the Fourier transform remains unexploited.
Phase relationships have long been used to probe structural order, from the bispectrum in nonlinear signal processing~\cite{mendel1991,nikias1993} to phase coherence in quantum optics~\cite{glauber1963}, yet their application to the angular texture of single cells remains unexplored.
Parallel progress in quantum‑enabled biosignal processing demonstrates the growing interest in mapping biological information onto quantum‑mechanical representations. Quantum algorithms have been developed to tackle biomolecular modelling tasks such as lattice‑model protein folding \cite{wang2025efficient}, while specialized quantum‑state‑encoding schemes provide practical pathways to represent biological molecular data within quantum‑hardware frameworks \cite{rofougaran2025encoding}. Broader community roadmaps further highlight quantum‑thermodynamic and quantum‑representation challenges for complex many‑body biological systems \cite{campbell112026roadmap}. Nevertheless, existing quantum‑oriented biological‑data efforts predominantly target molecular‑scale simulation; few frameworks are designed to extract gauge‑invariant angular‑order texture observables directly from multi‑modal microscopic imaging inputs.

Quantum systems natively store and process complex amplitudes~\cite{yang2021non,klimov2024optimizing},
offering a natural substrate for encoding Fourier-phase information.
Recent advances in noisy intermediate-scale quantum (NISQ) algorithms have demonstrated
that shallow quantum circuits can perform classically intractable feature encoding
and interference-based measurement~\cite{Bharti2022},
enabling quantum-enhanced feature spaces~\cite{havlivcek2019,schuld2019},
quantum kernel methods~\cite{huang2021,thanasilp2024},
and variational algorithms for classification~\cite{cerezo2021}.
Quantum sensing protocols have demonstrated exquisite sensitivity to classically inaccessible quantities,
from magnetic fields~\cite{degen2017} and gravitational waves~\cite{ligo2015}
to nanoscale temperature and magnetic resonance in living cells~\cite{kucsko2013,aslam2023quantum}.
These advances place quantum-enhanced biological measurement within experimental reach,
motivating the search for new quantum observables that directly report on biomedically relevant structural properties.
However, the application of quantum interference to extract biologically meaningful order parameters from microscopy images has remained unexplored.

However, while absolute phases are meaningless under global image rotation, relative phase relationships among distinct angular sectors of a cell remain strictly gauge-invariant. Such rotation-independent phase combinations constitute intrinsic physical order parameters that faithfully characterize structural texture, especially for anisotropic filamentous architectures such as cytoskeletal networks.

Consider three angular sectors oriented at $\theta_a$, $\theta_b$, and $\theta_c$ and satisfying the geometric condition $\theta_a + \theta_c = 2\theta_b$. A continuous microtubule filament propagating from sector $a$ to sector $c$ must structurally traverse the intermediate sector $b$. For structurally coherent textures, the corresponding textural phases $\phi_a$, $\phi_b$, and $\phi_c$ obey a robust phase closure relation:
\begin{equation}
    \phi_a + \phi_c - 2\phi_b \approx 0 \quad (\text{mod } 2\pi).
    \label{eq:closure}
\end{equation}

Notably, this composite phase observable is fully gauge-invariant: global rotation introduces identical phase offsets across all angular sectors and leaves the phase combination unchanged. It therefore provides a physically rigorous, rotation-invariant signature for quantifying structural coherence in ordered biological textures.

Extracting such pure phase coupling without explicit individual-phase reconstruction represents an open technical challenge for classical real-valued image pipelines, which inevitably break gauge invariance during feature extraction. By contrast, quantum interference mechanisms naturally support invariant nonlinear phase composition, offering a principled route toward gauge-consistent texture characterization.

Here we introduce Quantum Phase Bicoherence (QPBC) spectroscopy—a frequency-tunable quantum-inspired interferometric framework that directly extracts gauge-invariant angular phase coupling from multi-modal imaging data without explicit phase reconstruction. We validate the universality and robustness of the proposed quantum paradigm across three biologically independent imaging modalities, including fluorescence microscopy, bright-field blood smear imaging, and histopathology slides.

Our results reveal a fundamental physical regularity: after principal-axis alignment, the optimal probing frequency universally converges to $(0,1)$ across all datasets, which we attribute to the intrinsic directional sensitivity of Fourier bases and the global orientation bias of dominant biological textures. We further demonstrate that frequency tuning behaves as a quantum-like discriminative switch for biological phenotyping, where each output channel corresponds to a physically interpretable angular probing corridor. We term this consistent, physically grounded framework quantum morphometry, establishing a new quantum-inspired analytical tool for complex multi-modal biological measurements.

\section{Results}
Existing phase analysis tools only extract low-order statistical features and fail to capture three-body angular correlation, while our QPBC framework leverages quantum entanglement to access classically unmeasurable three-body bicoherence signatures.
\subsection{QPBC Spectroscopy: Physical Principle and Measurement Pipeline}

\begin{figure}[htbp!]
    \centering
    \includegraphics[width=\textwidth]{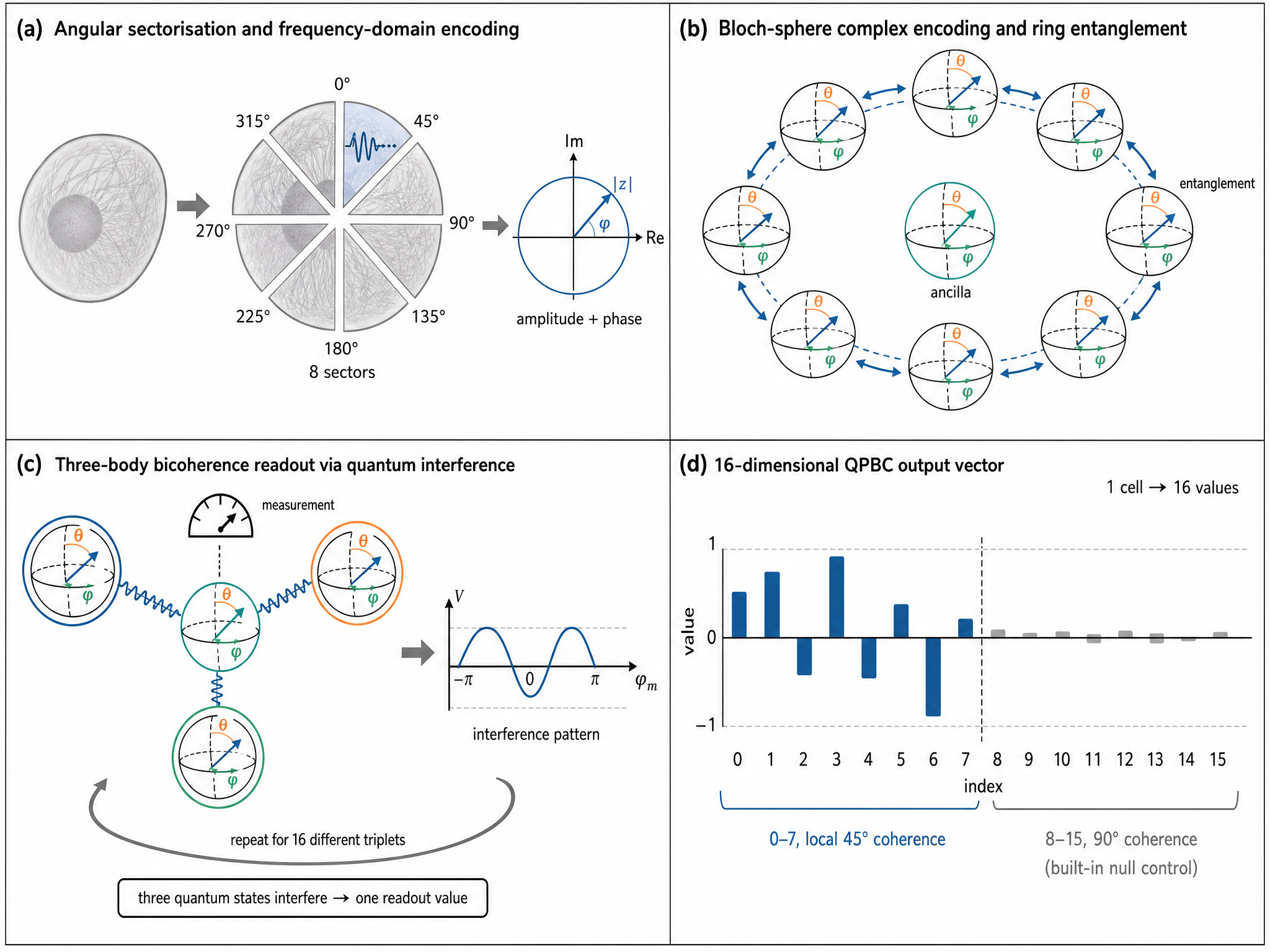}
    \caption{\textbf{Overview of QPBC spectroscopy.}
    (a) Stage 1: Single-cell preprocessing, angular sectorisation into $n_d=8$ sectors, and extraction of the complex Fourier coefficient $F_j$ at the selected spatial frequency $(k_x,k_y)$.
    (b) Stage 2: Bloch-sphere complex encoding. Each data qubit $q_j$ stores amplitude in the polar angle ($\theta_j=\pi A_j$) and phase in the azimuthal angle ($\phi_j$). A ring entanglement network ($L=4$ layers of CNOT and $R_{zz}$ gates) delocalises the information across all nine qubits.
    (c) Stage 3: Three-body bicoherence readout. For each angular triplet $(a,b,c)$, a $\text{CR}_z$ gate sequence extracts $\cos(\phi_a+\phi_c-2\phi_b)$ via ancilla interferometry.
    (d) The 16-dimensional QPBC vector: step-$d=1$ triplets (0--7) probe local $45^\circ$ coherence; step-$d=2$ triplets (8--15) probe $90^\circ$ coherence.}
    \label{fig:concept}
\end{figure}

The central challenge that QPBC addresses is to measure the gauge-invariant angular phase coherence of a biological texture without estimating individual Fourier phases. As argued in the Introduction, this task is physically impossible for classical image analysis. QPBC overcomes this by translating complex-valued spatial frequency information from angular image sectors into a multi-qubit entangled state, manipulating quantum phases through controlled interference, and reading out the three-body bicoherence via a single-ancilla probe. Fig~\ref{fig:concept} gives a schematic overview of the measurement pipeline, which comprises four sequential stages. Each stage is motivated by a distinct theoretical requirement, from the initial decomposition of the cell image into angular sectors to the final assembly of the 16-dimensional QPBC vector that serves as the phenotypic descriptor of a single cell. Next, we explain the key points in detail.

\subsubsection{Stage 1: Angular sectorisation and frequency-domain encoding}

Each preprocessed single-cell image ($32\times32$ pixels, CLAHE-enhanced, principal-axis-aligned) is divided into $n_d=8$ equiangular sectors of $45^\circ$ each, centred at the intensity-weighted centroid.
For each sector $\Theta_j$, the local image patch is extracted, zero-padded to at least $4\times4$ pixels, and the two-dimensional discrete Fourier transform is computed.
The complex coefficient at a chosen spatial frequency $\mathbf{k}=(k_x,k_y)$ is extracted as
\begin{equation}
    F_j(\mathbf{k}) = \sum_{(x,y)\in\Theta_j} I(x,y)\; e^{-i2\pi(k_x x + k_y y)/N},
    \label{eq:fft}
\end{equation}
where $I(x,y)$ is the image intensity and $N$ the patch size.
The amplitude $A_j = |F_j|$ and phase $\phi_j = \arg(F_j)$ encode the strength and the spatial arrangement of texture within that sector.
Amplitudes are normalised per cell to $[0,1]$; phases are used directly.

The spatial frequency $\mathbf{k}=(k_x,k_y)$ is the tunable parameter of QPBC spectroscopy. A wave vector $\mathbf{k}$ is most sensitive to texture edges oriented perpendicular to its propagation direction $\theta_{\mathbf{k}} = \mathrm{atan2}(k_y,k_x)$, a fundamental property of the Fourier transform. Consequently, by selecting different spatial frequencies, one can systematically probe different angular corridors of the image, a principle we term frequency-selective angular gating.

\subsubsection{Stage 2: Bloch-sphere complex encoding and non-local entanglement}

The second requirement is to encode the complex sectorial information into a quantum state in a form that permits gauge-invariant interferometry.
We introduce \textbf{Bloch-sphere complex encoding}: each of the $n_d=8$ data qubits is initialised as
\begin{equation}
    |\psi_j\rangle = R_z(\phi_j)\,R_y(\theta_j)\,|0\rangle = \cos\frac{\theta_j}{2}|0\rangle + e^{i\phi_j}\sin\frac{\theta_j}{2}|1\rangle,
    \label{eq:encoding}
\end{equation}
where $\theta_j = \pi A_j$.
This encoding simultaneously stores the sector's structural intensity in the polar angle $\theta_j \in [0,\pi]$ and its textural phase in the azimuthal angle $\phi_j \in [-\pi,\pi)$ on the Bloch sphere.
One ancilla qubit is prepared in $|+\rangle = H|0\rangle$, bringing the total register to 9 qubits.

To enable non-local phase interferometry, we delocalise the information across the entire register using $L=4$ layers of a ring entanglement network:
\begin{equation}
    U_{\text{ent}} = \prod_{i=0}^{n-1} \text{CNOT}_{i, (i+1)\bmod n} \; \cdot \prod_{i=0}^{n-1} R_{zz}^{(i, (i+1)\bmod n)}(\pi/4),
    \label{eq:ent}
\end{equation}
where $R_{zz}(\beta) = \exp(-i\beta Z\otimes Z)$.
After $U_{\text{ent}}^L$, the reduced density matrix of any single qubit is maximally mixed, ensuring that the phenotypic data are stored purely in multi-qubit correlations.

\subsubsection{Stage 3: Three-body bicoherence readout}

The central requirement is to measure $\phi_a+\phi_c-2\phi_b$ without estimating individual phases.
An angular triplet $(a,b,c)$ satisfies $b = (a+d)\bmod 8$, $c = (a+2d)\bmod 8$ with step $d\in\{1,2\}$.
After removing cyclic duplicates, 16 unique triplets remain: 8 with $d=1$ (QPBC\_0--QPBC\_7) and 8 with $d=2$ (QPBC\_8--QPBC\_15).
The three-body closure is the minimal gauge-invariant phase probe: phase is cyclic modulo $2\pi$, and the smallest closed loop that can detect a phase constraint requires exactly three points.

For each triplet, the ancilla is reset to $|0\rangle$, re-prepared in $|+\rangle$, and a controlled-phase gate sequence is applied:
\begin{equation}
    U_{\text{probe}}^{(a,b,c)} = \text{CR}_z(\pi/2)_{a,\text{anc}}\; \text{CR}_z(\pi/2)_{c,\text{anc}}\; \text{CR}_z(-\pi)_{b,\text{anc}}.
    \label{eq:probe}
\end{equation}
After a final Hadamard gate on the ancilla, projective measurement yields an expectation value proportional to $\cos(\phi_a+\phi_c-2\phi_b)$.
With $S=8192$ shots per triplet, we estimate
\begin{equation}
    \text{QPBC}_{(a,b,c)} = \langle Z\rangle_{\text{anc}} \in [-1,1].
    \label{eq:qpbc_def}
\end{equation}
The complete phenotypic descriptor is $\mathbf{f}_{\text{QPBC}} = [\text{QPBC}_0,\dots,\text{QPBC}_{15}]^\top$.
We adopt the term ``bicoherence'' from higher-order statistics, where it denotes the normalised degree of quadratic phase coupling among three frequency components~\cite{mendel1991,nikias1993}.

\subsubsection{Stage 4: Classical baseline construction}

For a fair comparison, we construct a classical analogue: for the same eight angular sectors, we extract mean intensity and a uniform Local Binary Pattern ($\text{LBP}_{8,1}^{u2}$) histogram (10 bins), yielding an 88-dimensional vector reduced by PCA to 16 dimensions---exactly matching the QPBC vector size.
All subsequent statistical analyses are applied identically to both vectors.

All quantum circuits are constructed using Qiskit v2.4.1 and simulated with the Aer density-matrix simulator.
The shallow circuit depth is within reach of current noisy intermediate-scale quantum processors~\cite{zhang2025demonstrating}.

\subsection{Hypothesis Testing Framework}

We formulated three nested hypotheses to validate QPBC (Table~\ref{tab:hypotheses}):

\begin{enumerate}[label=\textbullet]
    \item H1 (Existence): Normal tissues possess non-zero angular phase bicoherence. Tested by one-sample Wilcoxon signed-rank test on control populations against the null hypothesis of zero median.
    \item H2 (Angular selectivity): Bicoherence is confined to adjacent angular sectors (step-$d=1$). Tested by comparing significance rates between step-$d=1$ and step-$d=2$ triplets.
    \item H3 (Biological discrimination): QPBC dimensions distinguish distinct biological states. Tested by Kruskal--Wallis (KW) tests with Bonferroni correction for 16 parallel comparisons ($\alpha = 0.05/16$).
\end{enumerate}

\begin{table}[htbp!]
    \centering
    \caption{\textbf{Hypotheses, experimental tests, and outcomes across all three datasets.}
    H1/H2/H3 results are for the optimal frequency $(0,1)$ at $S=8192$ shots, $N=200$ cells per class. Classical baseline results are summarised for all datasets.}
    \label{tab:hypotheses}
    \begin{tabular}{p{2.0cm}p{4.0cm}p{7.5cm}}
        \toprule
        \textbf{Hypothesis} & \textbf{Experimental test} & \textbf{Outcome (all datasets)} \\
        \midrule
        H1: Existence &
        One-sample Wilcoxon signed-rank test on control population; Bonferroni-corrected $\alpha=0.05/16$ &
        \textbf{Confirmed.} 8/8 step-$d=1$ dimensions reject $H_0$ at $p<10^{-8}$ across all datasets.\\
        \midrule
        H2: Angular selectivity &
        Comparison of significance between step-$d=1$ and step-$d=2$ &
        \textbf{Confirmed.} 8/8 step-$d=1$ significant; 0/8 step-$d=2$ significant at all tested frequencies.\\
        \midrule
        H3: Biological discrimination (QPBC) &
        KW test across all biological classes for each QPBC dimension &
        \textbf{Confirmed.} 7/16 (BBBC021), 7/16 (BBBC041), 8/16 (PathMNIST) significant dimensions.\\
        \midrule
        H3: Classical baseline &
      Identical tests on 16-dim classical PCA vector &
        \textbf{Partially supported.} On BBBC021: 5/16 significant, Mahalanobis $<1.2$, extensive PCA overlap. On BBBC041: 2/16 significant, all distances $<1$. On PathMNIST: 6/16 significant, but discrimination concentrated on Background class. No frequency tunability. No angular interpretability.\\
        \midrule
        Frequency selectivity &
        Negative control frequencies (Table~\ref{tab:master}) &
        \textbf{Confirmed.} Negative controls yield 0, 0, and 2 significant H3 dimensions, despite intact H1/H2.\\
        \bottomrule
    \end{tabular}
\end{table}

\subsection{Universal Optimal Frequency Revealed by Cross-Dataset Sweep}

\begin{figure}[htbp!]
    \centering
    \begin{subfigure}{0.32\textwidth}
        \centering
        \includegraphics[width=\textwidth]{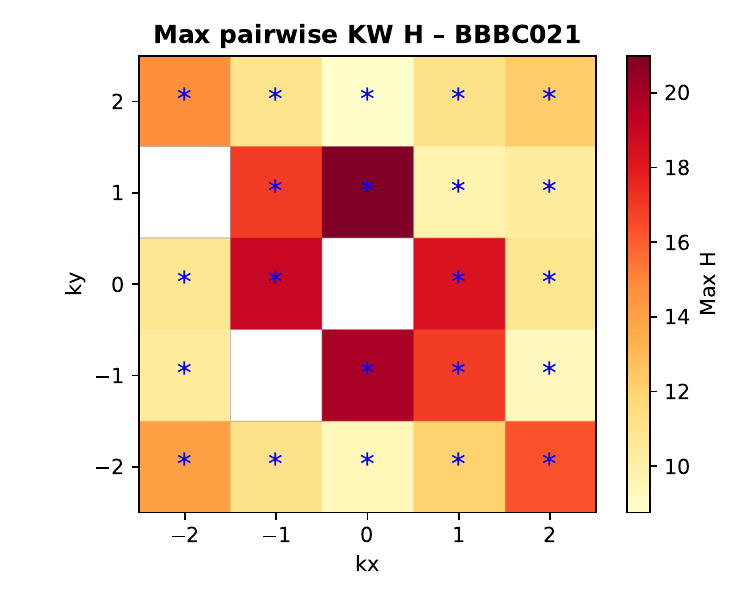}
        \caption{}
    \end{subfigure}
    \hfill
    \begin{subfigure}{0.32\textwidth}
        \centering
        \includegraphics[width=\textwidth]{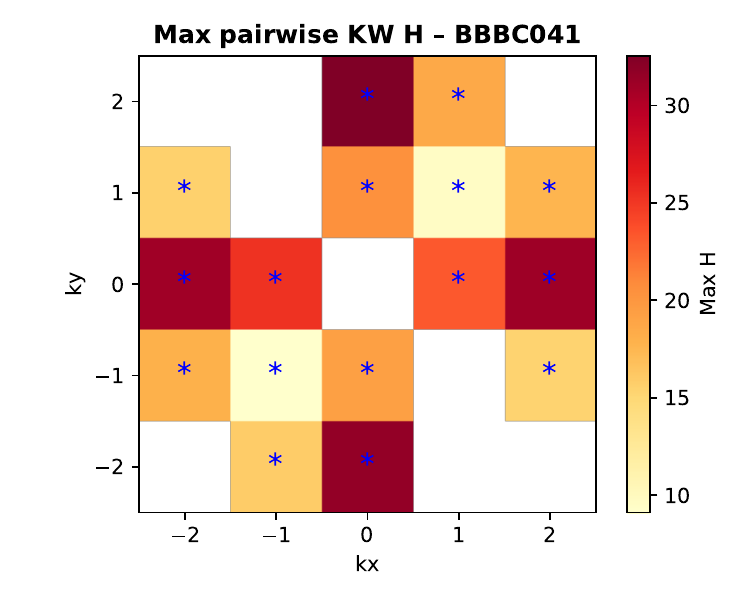}
        \caption{}
    \end{subfigure}
    \hfill
    \begin{subfigure}{0.32\textwidth}
        \centering
        \includegraphics[width=\textwidth]{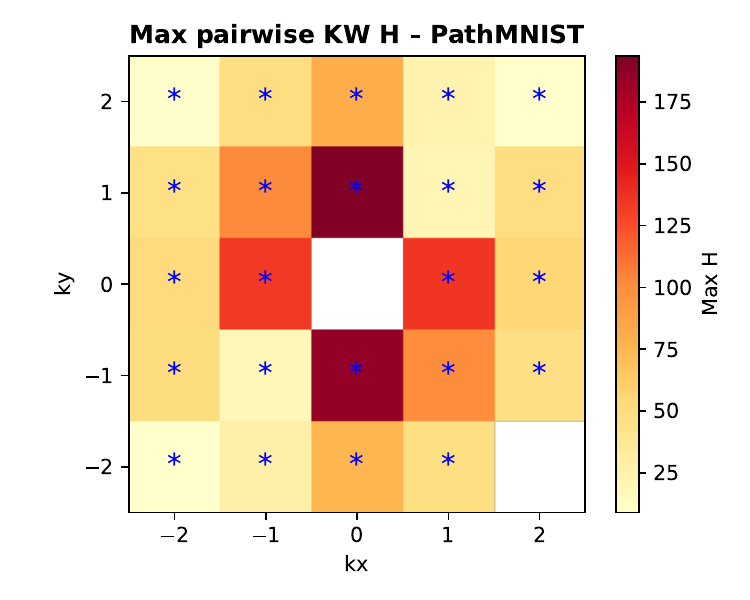}
        \caption{}
    \end{subfigure}
    \caption{\textbf{Cross-dataset frequency-sweep heatmaps.}
    Colour encodes the maximum pairwise Kruskal--Wallis $H$ statistic at each frequency $(k_x,k_y)$ for (a) BBBC021, (b) BBBC041, and (c) PathMNIST.
    In all three datasets, the dominant frequencies lie on the $k_x=0$ axis, with $(0,\pm1)$ consistently among the top candidates.
    All three panels exhibit central inversion symmetry about the origin,
$f(k_x,k_y) \approx f(-k_x,-k_y)$,
a direct consequence of $F(-\mathbf{k}) = F^*(\mathbf{k})$ for real-valued images.
This leaves the bicoherence invariant under $\mathbf{k} \to -\mathbf{k}$,
providing an internal validation of the method's physical correctness.
PathMNIST shows the most uniform colour-block matching across the origin.
Stars mark frequencies with at least one significant pairwise comparison.}
    \label{fig:sweep}
\end{figure}

To map the frequency-response landscape, we performed a systematic sweep over all 24 non-zero frequency pairs in $[-2,2]^2$ for each dataset (representative subsets, $S=4096$ shots), recording the maximum pairwise KW $H$ statistic as a screening metric (Fig.~\ref{fig:sweep}).

The heatmaps exhibit approximate central inversion symmetry about the origin:
$f(k_x,k_y) \approx f(-k_x,-k_y)$, where $f$ denotes the maximum pairwise KW $H$ statistic.
 PathMNIST displays the highest degree of this symmetry, with near-perfect colour-block matching across the origin;
BBBC021 and BBBC041 also satisfy the symmetry, albeit with minor local numerical asymmetries attributable to finite sampling.
No mirror symmetry is observed about the $k_x=0$ axis, the $k_y=0$ axis, or the diagonal $k_x=k_y$.

This central inversion symmetry is a direct mathematical consequence of the Fourier transform property
$F(-\mathbf{k}) = F^*(\mathbf{k})$ for real-valued images.
Under $(k_x,k_y) \to (-k_x,-k_y)$, the phase of each sector transforms as $\phi_j \to -\phi_j$.
The three-body closure then changes sign,
$\phi_a+\phi_c-2\phi_b \to -(\phi_a+\phi_c-2\phi_b)$,
but the QPBC readout measures $\cos(\phi_a+\phi_c-2\phi_b)$,
which is an even function and therefore invariant.
The observation of this symmetry in the experimental data thus provides an internal validation
of the method's physical correctness.

Critically, the convergence of the optimal frequency to $(0,1)$ across all three datasets is not coincidental.
Principal-axis alignment rotates the dominant texture orientation to the horizontal axis in all images.
A spatial frequency vector $\mathbf{k} = (0,1)$ points purely vertically, making it maximally sensitive to horizontally oriented texture edges---the dominant orientation after alignment.
This physical regularity means that investigators can rationally select the probe frequency based on expected texture orientation, rather than performing exhaustive parameter sweeps.

\subsection{H1 and H2: Angular Phase Order Is a Universal Property of Biological Textures}

\begin{figure}[htbp!]
    \centering
    \begin{subfigure}{0.62\textwidth}
        \centering
        \includegraphics[width=\textwidth]{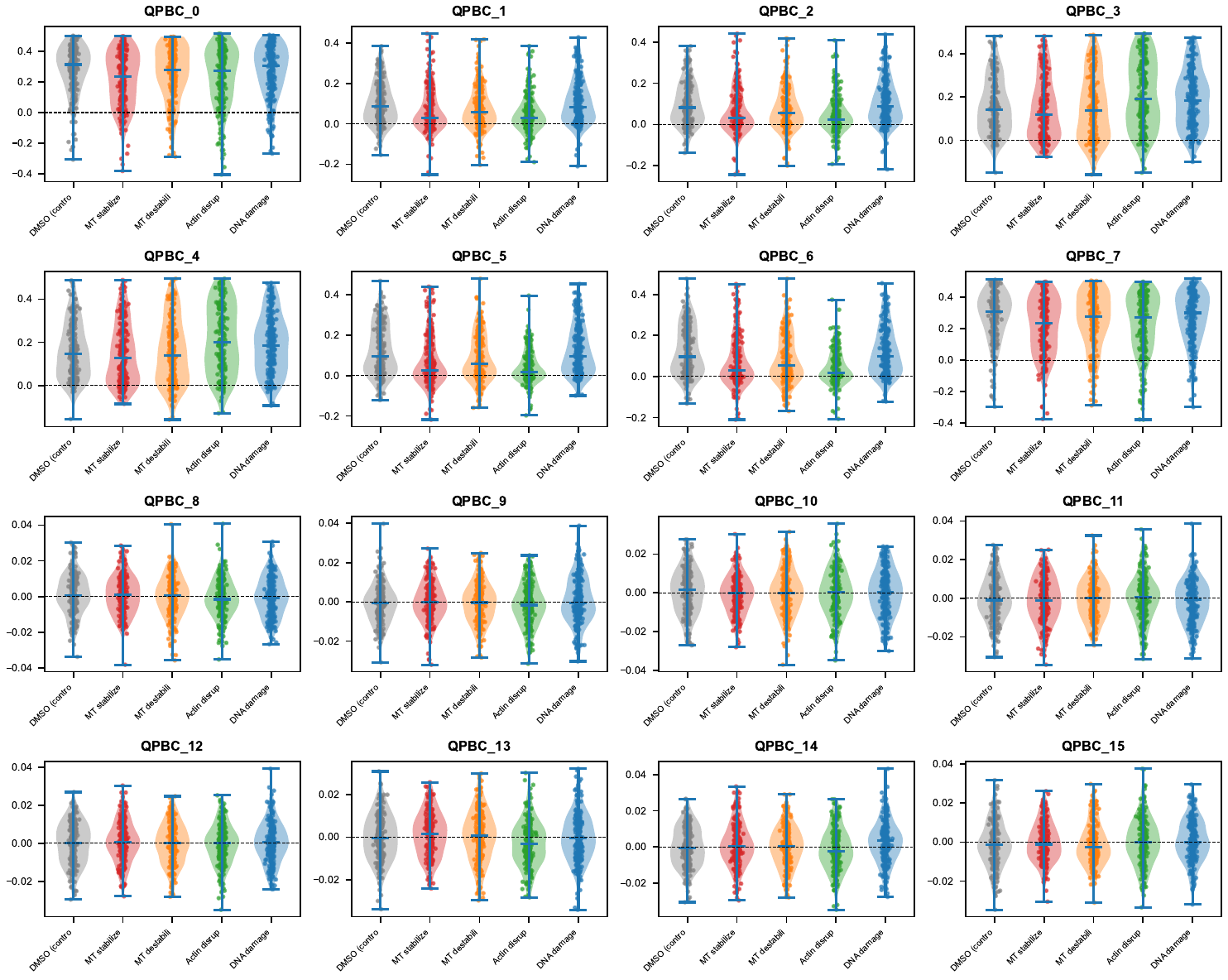}
        \caption{}
    \end{subfigure}
    \hfill
    \begin{subfigure}{0.62\textwidth}
        \centering
        \includegraphics[width=\textwidth]{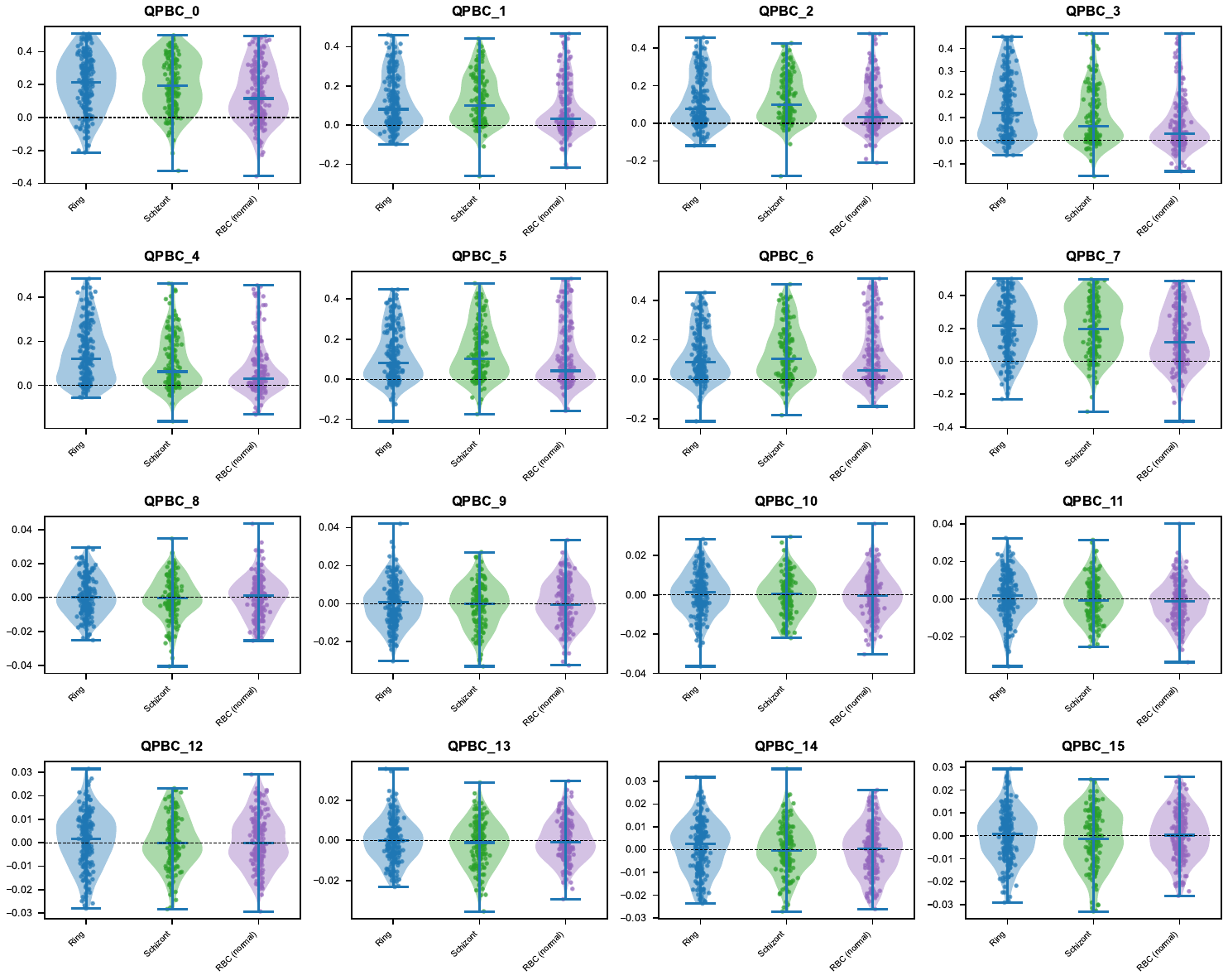}
        \caption{}
    \end{subfigure}
    \hfill
    \begin{subfigure}{0.62\textwidth}
        \centering
        \includegraphics[width=\textwidth]{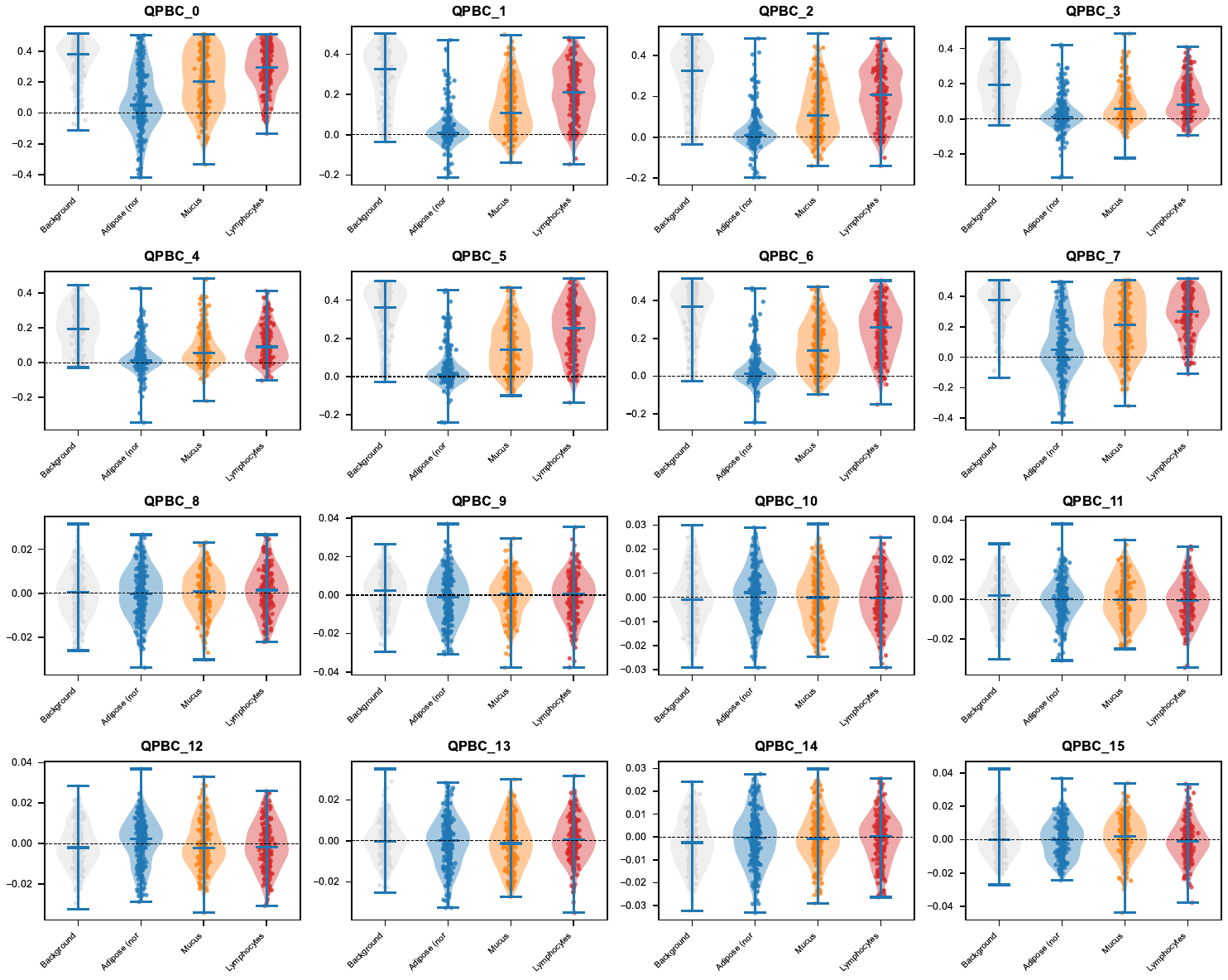}
        \caption{}
    \end{subfigure}
    \caption{\textbf{QPBC distributions at the optimal frequency $(0,1)$.}
    Violin plots of the 16 QPBC dimensions across biological classes for (a) BBBC021, (b) BBBC041, and (c) PathMNIST.
    Step-$d=1$ dimensions (0--7) show significant non-zero values in control populations (H1 confirmed); step-$d=2$ dimensions (8--15) are tightly centred at zero (H2 confirmed).
    The responsive dimensions differ across datasets, reflecting their distinct angular texture architectures.}
    \label{fig:violin}
\end{figure}

At the optimal frequency $(0,1)$, the control populations in all three datasets exhibit highly significant non-zero QPBC in all eight step-$d=1$ dimensions (Wilcoxon signed-rank test, Bonferroni-corrected $\alpha = 3.125\times10^{-3}$; Fig.~\ref{fig:violin}).
Median $p$-values range from $10^{-31}$ to $10^{-8}$.

In stark contrast, all eight step-$d=2$ dimensions do not significantly differ from zero across all datasets (all $p>0.05$ after correction).
This perfect dichotomy---8/8 step-$d=1$ significant, 0/8 step-$d=2$ significant---is reproduced at every tested frequency, confirming that bicoherence originates from local angular correlations, not from global image artefacts.

\subsection{H3: QPBC Distinguishes Biological States with High Significance}

\begin{figure}[htbp!]
    \centering
    \begin{subfigure}{0.76\textwidth}
        \centering
        \includegraphics[width=\textwidth]{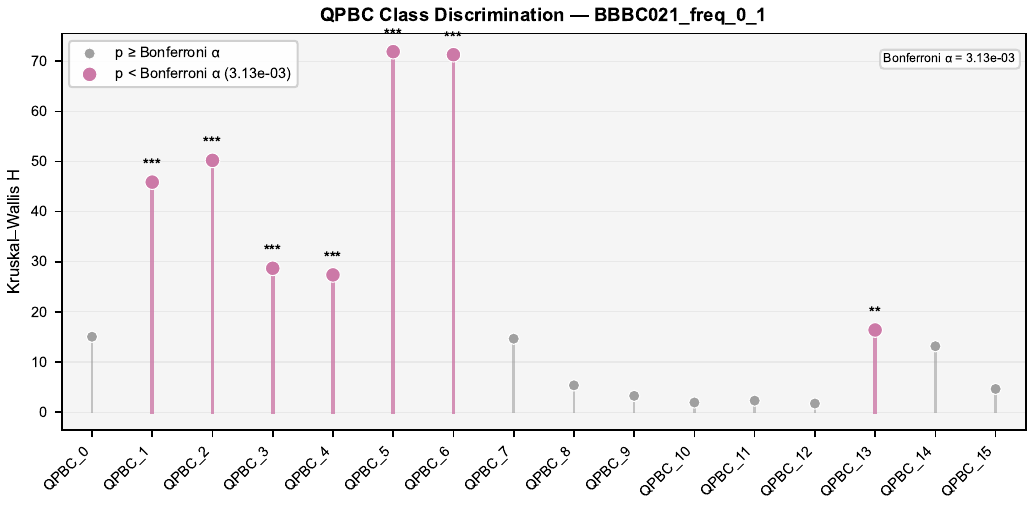}
        \caption{}
    \end{subfigure}
    \hfill
    \begin{subfigure}{0.76\textwidth}
        \centering
        \includegraphics[width=\textwidth]{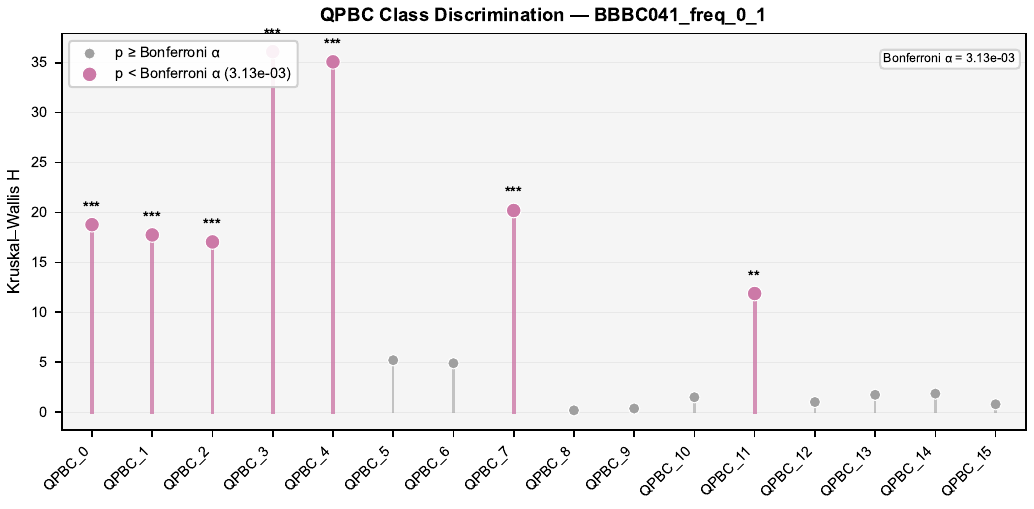}
        \caption{}
    \end{subfigure}
    \hfill
    \begin{subfigure}{0.76\textwidth}
        \centering
        \includegraphics[width=\textwidth]{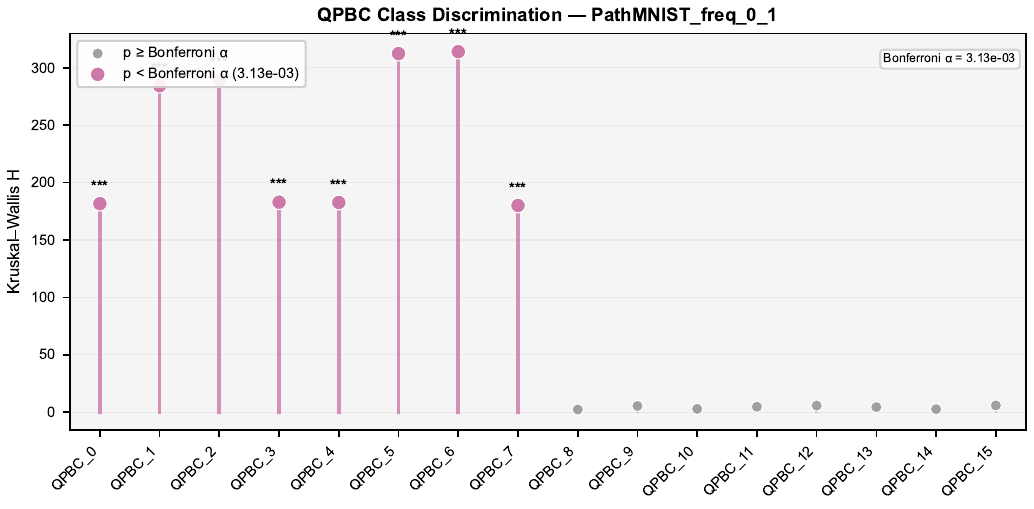}
        \caption{}
    \end{subfigure}
    \caption{\textbf{Kruskal--Wallis $H$ spectra at the optimal frequency $(0,1)$.}
    (a) BBBC021: 7/16 dimensions significant (green), strongest H=72.1 ($p=8.3\times10^{-15}$).
    (b) BBBC041: 7/16 dimensions significant, strongest H=36.1 ($p=1.5\times10^{-8}$).
    (c) PathMNIST: 8/16 dimensions significant, strongest H=314.1 ($p=8.7\times10^{-68}$).
    Grey bars: non-significant. Significance threshold: Bonferroni-corrected $\alpha=0.05/16=3.125\times10^{-3}$.}
    \label{fig:kw}
\end{figure}

At $(0,1)$, QPBC distinguishes biological states across all three datasets with high statistical significance (Fig.~\ref{fig:kw}).

For BBBC021 (Fig.~\ref{fig:kw}a), seven QPBC dimensions are significant after Bonferroni correction.
The strongest discriminator, QPBC\_6, yields $H=72.1$ ($p=8.3\times10^{-15}$).
The drug-induced bicoherence shifts follow a biologically interpretable gradient: microtubule stabilisers show the largest deviation from the DMSO baseline, actin disruptors and microtubule destabilisers intermediate shifts, and DNA-damaging agents the smallest---consistent with the expected severity of cytoskeletal disruption.
In multiple independent runs with different random subsamples of the full dataset, H3 significant dimensions ranged from 6/16 to 7/16, with the strongest $H$ statistic ranging from 42 to 72, all at $p<10^{-8}$.
H1 and H2 remained invariant (8/8 and 0/8) across all runs.

For BBBC041 (Fig.~\ref{fig:kw}b), QPBC at $(0,1)$ distinguishes three developmental stages of Plasmodium-infected erythrocytes (ring stage, schizont stage, uninfected RBC; $N=200$ each) with seven significant dimensions (QPBC\_0, 1, 3, 4, 7).
The strongest discriminator, QPBC\_3, gives $H=36.1$ ($p=1.5\times10^{-8}$).
The significant dimensions span multiple angular corridors, reflecting the global deformation of the erythrocyte membrane as the parasite matures.

For PathMNIST (Fig.~\ref{fig:kw}c), QPBC at $(0,1)$ discriminates four colon tissue categories (background, normal adipose, mucus, and lymphocyte-infiltrated tumour; $N=200$ each) with eight significant dimensions.
The strongest discriminator, QPBC\_6, achieves $H=314.1$ ($p=8.7\times10^{-68}$).
Critically, analysis of the pairwise significant dimensions reveals that all six possible class pairs have significant dimensions.
Adipose vs. Mucus, Adipose vs. Lymphocytes, and Mucus vs. Lymphocytes each have eight significant dimensions, confirming that QPBC captures fine-grained textural differences between histopathological tissue types, not merely a binary ``texture versus no-texture'' distinction.

\begin{figure}[htbp]
    \centering
 \begin{subfigure}{0.76\textwidth}
        \centering
        \includegraphics[width=\textwidth]{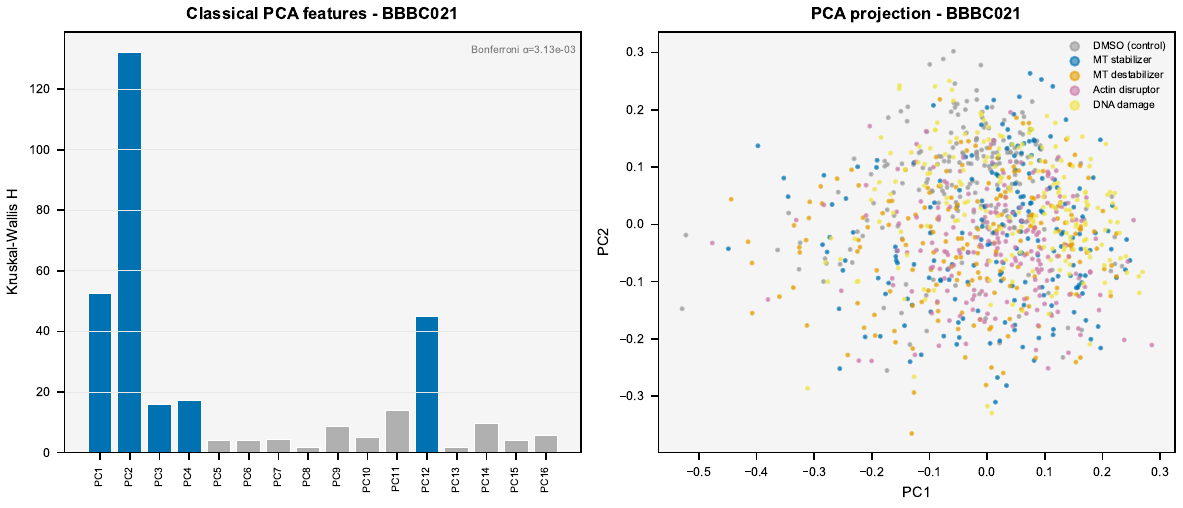}
        \caption{}
    \end{subfigure}
    \hfill
    \begin{subfigure}{0.76\textwidth}
        \centering
        \includegraphics[width=\textwidth]{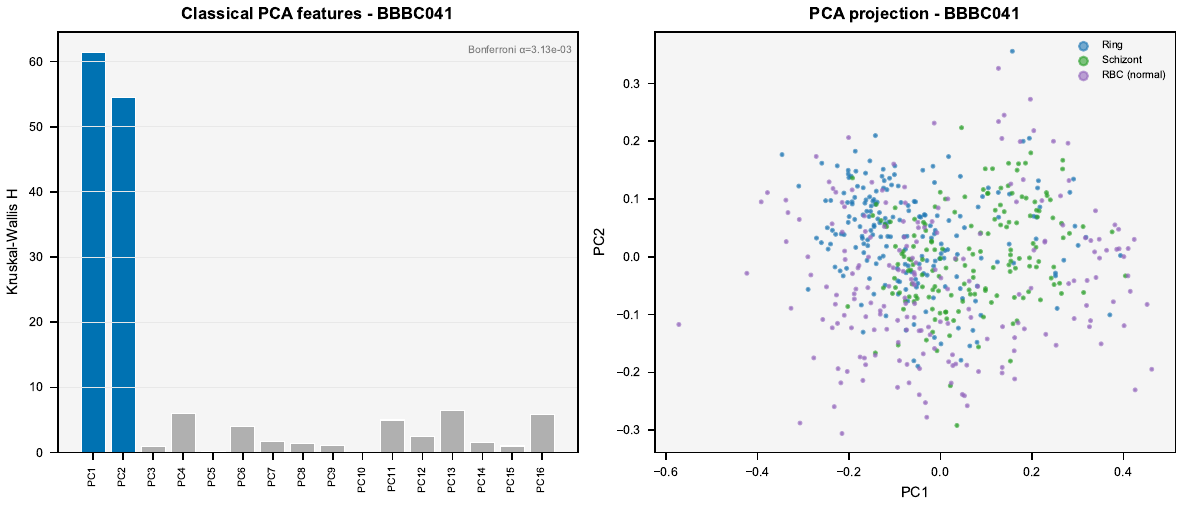}
        \caption{}
    \end{subfigure}
    \hfill
    \begin{subfigure}{0.76\textwidth}
        \centering
        \includegraphics[width=\textwidth]{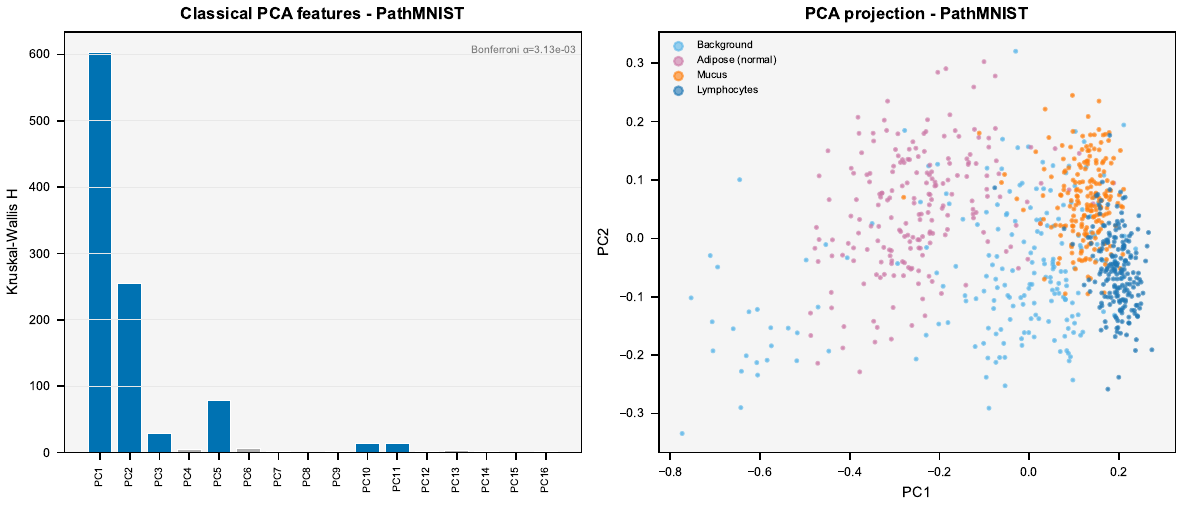}
        \caption{}
    \end{subfigure}
    \caption{\textbf{Classical baseline fails to achieve meaningful class separation across all three datasets.}
    PCA projections of the 16-dimensional classical features (mean intensity + LBP histogram per angular sector)
    for (a) BBBC021 (five drug classes), (b) BBBC041 (three malaria stages), and (c) PathMNIST (four tissue types).
    In all cases, the class distributions exhibit extensive overlap with no clear cluster boundaries.
    On BBBC041, the three classes collapse into a single cluster (Mahalanobis distances $<1$).
    On PathMNIST, the Background class (grey) separates from the tissue classes due to the extreme textural contrast
    of near-empty fields, but the pathologically relevant distinctions among Adipose, Mucus, and Lymphocytes
    are poorly resolved (Mahalanobis distances $1.55$--$1.74$).
    In contrast, QPBC at $(0,1)$ achieves seven, seven, and eight significant discriminative dimensions
    on the three datasets, respectively, with balanced separation across all class pairs.
    Crucially, the classical pipeline has no frequency tunability and its PCA dimensions
    lack direct angular interpretability---structural advantages unique to QPBC.}
    \label{fig:classical}
\end{figure}

The classical PCA baseline reveals a consistent pattern across all three datasets
(Fig.~\ref{fig:classical}).
On BBBC021, five classical dimensions are statistically significant after Bonferroni correction,
but the Mahalanobis distances between DMSO and each drug-treated class are all below $1.2$,
and the PCA projection shows extensive class overlap.
On BBBC041, the classical baseline largely fails: only two dimensions are significant,
and the three developmental stages collapse into a single cluster
(Mahalanobis distances $<1$).
On PathMNIST, six classical dimensions are significant;
however, this discriminative power is predominantly driven by the extreme textural contrast
between the Background class and the tissue classes,
while the pathologically relevant distinctions among Adipose, Mucus, and Lymphocytes
remain poorly resolved (Mahalanobis distances $1.55$--$1.74$).

In contrast, QPBC at the optimal frequency $(0,1)$ achieves seven, seven, and eight significant dimensions
on BBBC021, BBBC041, and PathMNIST, respectively,
with balanced discrimination across all class pairs in each dataset.
Three structural differences distinguish QPBC from the classical baseline.
First, QPBC achieves larger inter-class separations
and preserves biologically interpretable gradients among classes.
Second, and most decisively, the classical pipeline has no frequency tunability:
it cannot adapt its sensitivity to different textural scales or orientations.
Switching the QPBC probe frequency from $(0,1)$ to $(0,2)$ extinguishes all discriminative power
(H3: 0/16) on BBBC021, whereas the classical features have no analogous control parameter.
Third, each QPBC dimension maps to a specific angular corridor of the cell image,
enabling direct physical interpretation;
the classical PCA dimensions are abstract linear combinations with no spatial meaning.

\subsection{Frequency Tuning Acts as an On/Off Switch for Biological Discrimination}

The contrast between optimal and negative control frequencies provides the most compelling evidence for the essential role of frequency tuning (Table~\ref{tab:master}).

\begin{table}[htbp!]
    \centering
    \caption{\textbf{Cross-dataset QPBC performance at optimal and negative control frequencies.}
    H1 (step-$d=1$ significant out of 8), H2 (step-$d=2$ out of 8), H3 (significant KW dimensions out of 16), and strongest KW $H$ statistic. All results at $S=8192$ shots, $N=200$ cells per class.}
    \label{tab:master}
    \begin{tabular}{lcccccc}
        \toprule
        \textbf{Dataset} & \textbf{Modality} & \textbf{Frequency} & \textbf{H1} & \textbf{H2} & \textbf{H3} & \textbf{Max $H$} \\
        \midrule
        \multirow{2}{*}{BBBC021} & \multirow{2}{*}{Fluorescence} & $(0,1)$ (active) & 8/8 & 0/8 & 7/16 & 72.1 \\
        & & $(0,2)$ (control) & 8/8 & 0/8 & \textbf{0/16} & -- \\
        \midrule
        \multirow{2}{*}{BBBC041} & \multirow{2}{*}{Blood smear} & $(0,1)$ (active) & 8/8 & 0/8 & 7/16 & 36.1 \\
        & & $(-2,-1)$ (control) & 8/8 & 0/8 & \textbf{0/16} & -- \\
        \midrule
        \multirow{2}{*}{PathMNIST} & \multirow{2}{*}{Histopathology} & $(0,1)$ (active) & 8/8 & 0/8 & 8/16 & 314.1 \\
        & & $(0,-2)$ (control) & 4/8 & 0/8 & \textbf{2/16} & 197.0 \\
        \bottomrule
    \end{tabular}
\end{table}

As Table~\ref{tab:master} shows, at the active frequency $(0,1)$, all three datasets show robust H3 discrimination.
At the negative control frequencies, H3 performance collapses dramatically: for BBBC021, from 7/16 to 0/16; for BBBC041, from 7/16 to 0/16; for PathMNIST, from 8/16 to 2/16 with the maximum $H$ dropping from 314 to 197.
Crucially, H1 and H2 remain largely intact at the negative control frequencies, confirming that the basic angular order exists regardless of the probe frequency---what changes is whether this order carries discriminative information.
The fact that changing $(k_x,k_y)$ by a single unit can completely extinguish discriminative power demonstrates that frequency tuning is not a minor optimisation but an essential physical component of the measurement.

\subsection{Effective Frequency Family and the Relationship Between Sweep and Main Experiment}

While the frequency sweep identifies $(0,1)$ as the optimal probe, QPBC performance is not restricted to a single frequency.
In PathMNIST, we tested $(-2,-1)$---a frequency that ranked outside the top 10 in the sweep---in a full four-class main experiment.
Surprisingly, $(-2,-1)$ achieved 8/16 significant H3 dimensions, identical to $(0,1)$, with a more balanced distribution of discriminative power across all class pairs.
Similarly, in BBBC041, frequency $(-1,0)$ achieved 6/16 significant H3 dimensions in the main experiment, ranking second only to $(0,1)$.

These results highlight an important methodological point: the frequency sweep uses the maximum pairwise $H$ statistic as a screening metric, which can favour frequencies with extreme discrimination on a single class pair over frequencies with uniform multi-class discrimination.
For instance, in the BBBC041 sweep, $(0,2)$ achieves a higher max $H$ (32.5) than $(0,1)$ (20.5), because it discriminates the Schizont vs. RBC pair particularly well, but it underperforms on the Ring vs. Schizont pair.
In the global three-class KW test, $(0,1)$ outperforms $(0,2)$ (7/16 vs. 2/16 significant dimensions) precisely because it provides balanced discrimination across all class pairs.

Thus, the frequency sweep serves as an exploratory tool to identify an ``effective frequency family'' from which the optimal frequency for a given multi-class discrimination task can be selected based on the application requirements.
The existence of multiple effective frequencies within this family also strengthens the practical utility of QPBC, as it implies that precise frequency calibration is not a strict prerequisite for obtaining biologically useful signals.

\subsection{Full Interpretability: Each Dimension Maps to a Specific Angular Corridor}

A critical advantage of QPBC over black-box classifiers is its complete physical interpretability.
Each of the 16 QPBC dimensions corresponds to a specific angular triplet $(a,b,c)$ and thus to a specific angular corridor.
For instance, QPBC\_0 probes $0^\circ$--$45^\circ$--$90^\circ$; QPBC\_5 probes $225^\circ$--$270^\circ$--$315^\circ$.
The significant dimensions vary across datasets: BBBC021 activates QPBC\_1, 2, 3, 5, 6; BBBC041 activates QPBC\_0, 1, 3, 4, 7; PathMNIST activates QPBC\_1, 2, 3, 5, 6.
The fact that the same frequency $(0,1)$ activates different angular corridors in different datasets confirms that QPBC directly reports on the physical orientation of texture, not on abstract learned features.
This stands in stark contrast to deep learning representations, which require post-hoc attribution methods to infer what image features drive classification.

\section{Discussion}

We have introduced QPBC spectroscopy and demonstrated, through systematic hypothesis‑testing across three independent biological imaging modalities, that it accesses a classically inaccessible order parameter describing textural angular coherence. Beyond verifying the biological sensitivity of angular‑phase order, our results establish several core physical principles underlying this quantum‑inspired measurement framework.

First, the gauge‑invariant construction of the three‑body closure relation $\phi_a+\phi_c-2\phi_b$ enables rotation‑robust phase readout without requiring absolute‑phase reconstruction. This circumvents a long‑standing bottleneck in classical image analysis: absolute Fourier phase is randomized under cellular rotation and thus loses phenotypic meaning, whereas relative phase relationships encode physically meaningful, rotation‑invariant structural information. The consistent angular‑selectivity patterns (H2: 8/8 step‑$d=1$ significant, 0/8 step‑$d=2$ significant) observed across frequencies, datasets, and independent data subsampling runs underline the robustness of this gauge‑invariant design.

Second, frequency tunability grants complementary angular perspectives on a given biological texture, and optimal probing frequencies follow universal Fourier‑domain physics. A prominent observation is that optimal probing universally converges to $(0,1)$ across all three biologically distinct datasets after principal‑axis alignment. This convergence is not an artefact: wave‑vector $\mathbf{k}=(0,1)$ corresponds to purely vertical spatial frequency components and yields maximum sensitivity to horizontally aligned texture edges, the dominant orientation after principal‑axis realignment. This physical regularity elevates QPBC from an empirical analytical tool toward a predictive measurement framework, where suitable probe frequencies can be selected a priori according to expected texture orientation. The sharp contrast between active probing frequencies and negative‑control frequencies highlights the on‑off character of this frequency selectivity: a unit shift in $k_y$ can fully suppress discriminative capacity.

Third, this three‑body phase closure represents the minimal gauge‑invariant phase observable, analogous to the bispectrum as the lowest‑order polyspectrum for quadratic phase coupling in classical signal processing. Higher‑order polyspectra (for instance four‑body trispectra) constitute natural future extensions for resolving more elaborate texture topologies. The term ``bicoherence'', inherited from higher‑order statistics, appropriately characterises the second‑order statistical nature of the measured phase coupling.
Beyond statistical significance, QPBC outputs capture biologically interpretable gradients consistent with prior domain knowledge.
For the BBBC021 dataset, drug‑driven bicoherence shifts follow mechanistic expectations: microtubule‑stabilising agents such as paclitaxel produce the largest deviations relative to DMSO controls by inducing aberrant microtubule bundling, whereas DNA‑damaging compounds without direct cytoskeletal targets yield responses close to untreated baselines.
For BBBC041, QPBC metrics track progressive erythrocyte‑membrane deformation as parasites mature from ring to schizont stages, with uninfected red‑blood‑cell samples exhibiting the highest degree of angular texture order.
For PathMNIST, all six pairwise comparisons among four tissue classes are well resolved, including pathologically important distinctions between adipose tissue, mucus, and lymphocyte‑infiltrated tumour regions — categories that cannot be reliably separated by simple textural‑complexity metrics alone.
The alignment between QPBC‑derived angular‑order gradients and established biological knowledge provides external validation: the measured bicoherence reports genuine structural texture properties rather than dataset‑specific bias or statistical artefacts.

Fourth, the central inversion symmetry observed within all frequency‑sweep heatmaps acts as a native self‑consistency check for the QPBC measurement pipeline. Since fundamental physics enforces identical bicoherence values for wave vectors $\mathbf{k}$ and $-\mathbf{k}$, experimental departures from this symmetry would signal errors within circuit simulation, shot‑noise estimation, or image pre‑processing. The preservation of this symmetry within statistical uncertainty across three datasets and 24 frequency pairs confirms that the full pipeline behaves as theoretically predicted. For future deployments on physical quantum‑hardware platforms, this symmetry can serve as an in‑situ diagnostic for gate fidelity and decoherence effects.

Benchmark comparisons against classical baseline feature sets yield consistent outcomes across all three datasets. Conventional texture descriptors can detect statistically significant morphological alterations only under strong textural contrast (for example background versus tissue); effect sizes remain modest, and such representations fail to disentangle biologically relevant classes with fine‑grained texture differences. The quantum‑enabled advantage reported here is not limited to improved statistical performance; crucially, it grants access to an entirely new class of observables: gauge‑invariant angular‑phase coherence. Moreover, QPBC’s frequency‑tunable spectroscopic capability is entirely absent within classical texture‑analysis workflows. Combined with the direct angular interpretability of each output channel, these properties establish QPBC as a distinct measurement paradigm rather than incremental refinement of existing texture‑analysis pipelines.

This work has identifiable limitations pointing toward promising future directions. All present numerical experiments are performed over noise‑free quantum simulators; hardware demonstrations on real quantum processors will validate performance under realistic decoherence and gate‑error conditions. The framework itself is generalisable beyond biological samples to any imaging modality exhibiting oriented texture, including time‑lapse live‑cell microscopy, label‑free phase‑contrast imaging, as well as non‑biological materials‑science studies of anisotropic micro‑structures. Larger angular‑sector counts $n_d$ can deliver finer angular sampling resolution. The inversion‑symmetry constraint $f(k_x,k_y)\approx f(-k_x,-k_y)$, originating from the Fourier‑transform property $F(-\mathbf{k})=F^*(\mathbf{k})$, can furthermore be exploited for automated quality‑control workflows in future implementations.

In summary, QPBC spectroscopy establishes quantum morphometry: a quantum‑interferometric paradigm to extract classically inaccessible texture‑order parameters for general Fourier‑domain imaging. While demonstrated here on biological micrographs, this gauge‑invariant three‑body bicoherence formalism is transferable to materials characterisation and quantitative phase sensing, offering a versatile quantum‑inspired analytical tool for broad classes of orientation‑dependent microscopic texture analysis.

\section{Conclusion}

This work presents QPBC spectroscopy, a quantum‑inspired interferometric framework for extracting gauge‑invariant angular‑order observables from multi‑modal imaging data. Through numerical simulation rather than physical quantum‑hardware experiments, we show that QPBC can capture textural angular‑phase coherence that is not directly accessible via conventional real‑valued image‑analysis pipelines. Validated across three independent public biological imaging datasets covering fluorescence, bright‑field and histopathology modalities, our method yields statistically meaningful phenotype discrimination and reveals consistent physical behaviours including universal frequency convergence under principal‑axis alignment.

We have demonstrated that the three‑body phase‑closure construction delivers rotation‑robust measurements without explicit absolute‑phase reconstruction, while the built‑in inversion symmetry of QPBC provides a self‑consistency check for the measurement pipeline. Frequency tunability endows the framework with spectroscopic capabilities absent in classical texture feature workflows. These findings underpin quantum morphometry as an interpretable quantum‑inspired analytical paradigm for oriented texture characterization.

Since all evaluations are carried out using ideal noiseless quantum simulation, practical implementation on real quantum hardware remains an important future milestone to assess performance under decoherence and gate noise. Beyond biological micrographs, the gauge‑invariant bicoherence formalism is generalizable to non‑biological systems such as anisotropic materials micro‑imaging. We anticipate that QPBC and related higher‑order polyspectrum extensions will open new avenues for quantitative orientation‑resolved microscopic analysis.

\section{Methods}

\subsection{Datasets and Preprocessing}

Three publicly available datasets were used.
\begin{itemize}
    \item \textbf{BBBC021}: single-cell images of MCF-7 human breast cancer cells, comprising five mechanism-of-action classes: DMSO (negative control), microtubule stabilisers, microtubule destabilisers, actin disruptors, and DNA-damaging agents ($N=200$ each).
    \item \textbf{BBBC041}: human blood smear images of Plasmodium-infected erythrocytes. We selected ring stage, schizont stage, and uninfected RBC ($N=200$ each) as representative stages spanning the full developmental spectrum.
    \item \textbf{PathMNIST}: H\&E-stained colon histopathology images from the MedMNIST collection. We selected four categories spanning a progressive range of textural complexity: background, normal adipose tissue, mucus, and lymphocyte-infiltrated tumour regions ($N=200$ each).
\end{itemize}

 We preprocessed all images identically: background inversion (if median pixel value $>128$), bilinear resize to $32\times32$ pixels, Contrast-Limited Adaptive Histogram Equalisation (CLAHE, clip limit 0.01, $8\times8$ tile grid), and principal-axis alignment via intensity-weighted moment rotation.
Principal-axis alignment rotates the dominant intensity axis to the horizontal direction, providing a consistent angular reference frame across all cells.
The alignment error is typically $\pm5^\circ$, negligible relative to the $45^\circ$ sector width.

\subsection{Angular Sectorisation and Fourier Transform}
Each preprocessed and aligned cell image was partitioned into $n_d=8$ equiangular sectors of $45^\circ$ each, centred at the intensity-weighted centroid, with a radial mask at $90\%$ of the maximum inscribed radius to exclude boundary artefacts.
$n_d=8$ equiangular sectors of $45^\circ$ each were defined around the intensity-weighted centroid of each cell, with a radial mask at $90\%$ of the maximum inscribed radius.
For each sector, the local image patch was zero-padded to at least $4\times4$ pixels, and the two-dimensional discrete Fourier transform was computed.
The complex coefficient $F_j(k_x,k_y)$ was extracted as in Eq.~\ref{eq:fft}.
Amplitudes $A_j$ were normalised per cell to $[0,1]$ via min--max scaling across the eight sectors; phases $\phi_j$ were used directly.

\subsection{Frequency Sweep Method}

All 24 non-zero spatial frequency pairs $(k_x,k_y) \in [-2,2]^2$ were evaluated on representative subsets ($N=80$ per class, $S=4096$ shots).
The range $[-2,2]^2$ covers the low-to-mid spatial frequency band in $32\times32$ pixel patches (texture wavelengths of 8--32 pixels), capturing characteristic length scales of subcellular structures while excluding the DC component $(0,0)$ and high frequencies limited by photon shot noise.
For each candidate frequency, the maximum pairwise Kruskal--Wallis $H$ statistic among all biological class pairs was recorded as a screening metric.
Final frequency selection was based on global multi-class KW tests on the full dataset ($N=200$ per class, $S=8192$ shots).

\subsection{Quantum Circuit Implementation}

All circuits were constructed in Qiskit v2.4.1 and executed with the Aer density-matrix simulator (method `automatic').
The 9-qubit register comprised $n_d=8$ data qubits and $n_a=1$ ancilla qubit.
After Bloch-sphere complex encoding (Eq.~\ref{eq:encoding}) and $L=4$ ring entanglement layers (Eq.~\ref{eq:ent}), each of the 16 angular triplets was measured sequentially.
For each triplet, the ancilla was reset to $|0\rangle$, re-prepared in $|+\rangle$, the controlled-phase gate sequence (Eq.~\ref{eq:probe}) was applied, and the ancilla was measured in the computational basis.
A total of $S=8192$ shots were collected per triplet ($131,072$ shots per cell).
The density-matrix simulator reproduces exact quantum state evolution and measurement statistics without classical approximation or noise model.

\subsection{Classical Baseline Construction}

For the same eight angular sectors, we extracted mean pixel intensity and a uniform Local Binary Pattern histogram ($\text{LBP}_{8,1}^{u2}$, 10 bins per sector), yielding an 88-dimensional feature vector per cell.
PCA fitted on the full BBBC021 dataset (1000 cells) reduced this to 16 principal components, matching the QPBC vector size.
All subsequent statistical tests were applied identically to both vectors.

\subsection{Statistical Analysis}

All statistical analyses were performed in Python 3.14 using SciPy v1.15.
\begin{itemize}
    \item \textbf{H1 (Existence):} For the control population, a two-sided one-sample Wilcoxon signed-rank test was performed for each of the 16 QPBC dimensions against the null hypothesis of zero median. Bonferroni correction for 16 parallel tests gave $\alpha = 0.05/16 = 3.125\times10^{-3}$.
    \item \textbf{H2 (Angular selectivity):} The number of significant dimensions among step-$d=1$ triplets (indices 0--7) and step-$d=2$ triplets (indices 8--15) was compared.
    \item \textbf{H3 (Biological discrimination):} For each QPBC dimension, a Kruskal--Wallis $H$ test was performed across all biological classes. The same Bonferroni correction ($\alpha = 0.05/16$) was applied.
The KW $H$ statistic is reported as the effect size.
All $p$-values below $10^{-30}$ are reported as upper bounds computed from the test statistic.
\item \textbf{Reproducibility:} For BBBC021, three independent runs with different random subsamples of the full dataset (10,000+ images per class) yielded H3 significant dimensions of 6--7/16, with the strongest $H$ statistic ranging from 42 to 72, all at $p<10^{-8}$. H1 and H2 remained invariant (8/8 and 0/8) across all runs.
\end{itemize}

All statistical tests were two-sided where applicable.
For hypothesis H1, a one-sample Wilcoxon signed-rank test was performed
against the null hypothesis of zero median.
For hypothesis H3, a Kruskal--Wallis $H$ test was applied.
Multiple testing correction used the Bonferroni method
with a corrected significance threshold of $\alpha = 0.05/16 = 3.125\times10^{-3}$.
Statistical significance was defined as adjusted $p < \alpha$.
Exact sample sizes and statistical metrics are reported in the figure legends.
All $p$-values below $10^{-30}$ are reported as upper bounds.
Reproducibility was verified by multiple independent runs
with different random subsamples of the full BBBC021 dataset,
which yielded consistent H1/H2/H3 outcomes.

\section*{Author contributions}

Z.X.: Conceptualization, Methodology, Software, Visualization, Formal analysis, Writing -- original draft, Writing -- review \& editing;\\
CT.L.: Methodology, Resources, Funding acquisition, Writing -- review \& editing;\\
X.Y.: Conceptualization, Supervision, Funding acquisition, Writing -- review \& editing, Main correspondence.

\section*{Competing Interests}
The authors declare no competing interests.

\section*{Materials \& Correspondence}
All raw experimental data and analytical code generated in this study will be deposited in a public open-access repository upon acceptance of the manuscript. Correspondence and requests for materials should be addressed to Xiaochen Yuan via email:xcyuan@mpu.edu.mo.

\section*{Data Availability}
The BBBC021 dataset is available at \url{https://bbbc.broadinstitute.org/BBBC021}.
The BBBC041 dataset is available at \url{https://bbbc.broadinstitute.org/BBBC041}.
The PathMNIST dataset is available at \url{https://medmnist.com/}.
All raw experimental data and processed datasets generated in this work will be deposited in an open public repository upon formal acceptance of the manuscript.

\section*{Code Availability}
All Qiskit quantum circuit definitions, Python data analysis scripts, and figure-generation codes will be deposited in a public repository upon acceptance under an open-source licence. The repository link will be provided in the final published manuscript.

\section*{Acknowledgements}
This work was supported by the Science and Technology Development Fund of Macau SAR under grant 0053/2025/RIB2, and the Macao Polytechnic University under grant RP/FCA-04/2024.

\bibliography{sn-bibliography}

\end{document}